%% file: revision.tex
\documentclass[superscriptaddress,twocolumn,amsmath,amssymb,aps,prc]{revtex4-2}
\usepackage{amsmath}
\usepackage{braket}
\usepackage{booktabs}
\usepackage{multirow}
\usepackage{physics}
\usepackage{xcolor}
\usepackage{framed}
\usepackage[normalem]{ulem}
\usepackage[dvipsnames]{xcolor}
\usepackage{graphicx}
\usepackage{dcolumn}
\usepackage{multirow}
\usepackage{longtable}
\usepackage{bm}
\usepackage{float}
\usepackage{hyperref}
\hypersetup{
    colorlinks=true,
    linkcolor=blue,
    filecolor=blue,
    citecolor=blue,
    urlcolor=blue,
    pdftitle={Overleaf Example},
    pdfpagemode=FullScreen,
    pdfauthor=author
    }
\begin{document}

\preprint{APS / Physical Review C}
\title{Determining the dynamic
deformation of $^{140}$Ce by constraining coupled-channels
parameters for fusion}
\author{Chandra Kumar}
\altaffiliation{Present address: Govt. T.C.L. P.G. College, Khokhra Bhata,
Janjgir 495668, Chhattisgarh, India}
\affiliation{Nuclear Physics Group, Inter-University Accelerator Centre,
Aruna Asaf Ali Marg, New Delhi 110067, India}
\author{Rohan Biswas}
\altaffiliation{Present address: Life Science Division, Diamond Light Source
Ltd., Diamond House, Harwell Science and Innovation Campus,
Didcot, Oxfordshire, OX11 0DE, United Kingdom}
\affiliation{Nuclear Physics Group, Inter-University Accelerator Centre,
Aruna Asaf Ali Marg, New Delhi 110067, India}
\author{J. Gehlot} 
\affiliation{Nuclear Physics Group, Inter-University Accelerator Centre,
Aruna Asaf Ali Marg, New Delhi 110067, India}
\author{Gonika}
\affiliation{Nuclear Physics Group, Inter-University Accelerator Centre,
Aruna Asaf Ali Marg, New Delhi 110067, India}
\author{A. Parihari}
\altaffiliation{Present address: Department of Physics, Rajdhani College,
University of Delhi, Mahatma Gandhi Marg, Raja Garden, New Delhi 110015}
\affiliation{Department of Physics and Astrophysics, Delhi University,
Delhi 110007, India}
\author{N. Madhavan}
\affiliation{Nuclear Physics Group, Inter-University Accelerator Centre,
Aruna Asaf Ali Marg, New Delhi 110067, India}
\author{A. Vinayak}
\affiliation{Department of Physics, Karnatak University, Dharwad 580003,
India}
\author{Amritraj Mahato}
\affiliation{Department of Physics, Central University of Jharkhand,
Ranchi 835205, India}
\author{S. Nath}                   
\email{subir@iuac.res.in}
\affiliation{Nuclear Physics Group, Inter-University Accelerator Centre,
Aruna Asaf Ali Marg, New Delhi 110067, India}
\date{\today}

\begin{abstract}
We present a systematic study of the
dynamic deformation of
$^{140}$Ce using $^{16}$O and $^{36}$S projectiles in heavy-ion
fusion reactions, combining experimental data,
a Gaussian analytic-barrier framework and
coupled-channels calculations. Fusion cross sections for
$^{16}$O+$^{140}$Ce are measured from $\simeq$17\% above to
$\simeq$12.4\% below the Bass barrier. Fusion data for
$^{36}$S+$^{140}$Ce are obtained from the literature. Deformation
parameters of $^{140}$Ce are extracted via chi-square minimization
and Bayesian analysis, with iNdependent Bayesian Model Averaging
yielding $\beta_2 = 0.09 \pm 0.03$ and $\beta_3 = 0.18 \pm 0.02$,
consistent across both systems. The extracted parameters are tested
in the $^{28}$Si+$^{140}$Ce system, where coupled-channels calculations
including transfer of a pair of neutrons ($2n$) reproduce both the
fusion excitation function and the barrier distribution. The positive
$Q$-value $2n$-pickup channel enhances fusion in this reaction,
while the projectile’s vibrational or rotational nature results
in similar structure of the barrier distribution. This study
demonstrates that the Gaussian analytic recipe is quite effective
in deriving the fusion barrier distribution which proves to be a
sensitive probe of intrinsic nuclear deformation. Further,
coupled-channels analysis across multiple systems ensures
robustness of the extracted deformation parameters.
\end{abstract}
\maketitle

Understanding femtometer-scale nuclear shapes, which arise from strong
nucleon correlations, remains a central challenge in nuclear physics
\cite{David2025}. To characterize these intrinsic shapes, nuclear shape
parameters \textendash such as quadrupole ($\beta_2$), octupole ($\beta_3$)
and hexadecapole ($\beta_4$) deformations \textendash have been introduced.
Their extraction relies on a combination of experimental investigations and
theoretical modeling. Data from both low-energy nuclear reactions
(see Ref.~\cite{Chandra2025PLB} and references therein) and high-energy
collisions, together with theoretical approaches, are widely employed
to probe nuclear deformation.

At relativistic energies,
key observables in heavy-ion collisions include
collective flow patterns and particle momentum distributions~\cite{STAR2025},
which provide insights into nuclear deformation. Though, a recent article
\cite{Dobaczewski2025} has raised important questions about the physical
foundation of extracting nuclear shape information from ultrarelativistic
heavy-ion collisions. At lower
energies, reaction dynamics \textendash particularly scattering and fusion
processes \textendash serve as sensitive probes of deformation effects
when interpreted within theoretical frameworks such as the coupled-channels
(CC)~\cite{Hagino2012,Hagino1999} and coupled reaction channel (CRC)
methods~\cite{Thompson1988}. For instance, in heavy-ion fusion reactions,
experimentally measured fusion cross sections ($\sigma_{\mathrm{fus}}$)
often exceed the predictions of single-barrier penetration models near
and below the Coulomb barrier. This enhancement can be understood in terms
of nuclear deformation effects~\cite{Stokstad1978} in many systems,
whereby CC approaches explicitly couple the relative motion of the
colliding nuclei to their intrinsic degrees of freedom. The CC framework
successfully reproduced a broad range of experimental fusion data
\cite{Dasgupta1998,Back2014,Montagnoli2017,Montagnoli2023}.
Consequently, fusion reactions have been found to be very useful in
the study of nuclear shapes
and deformation parameters have been extracted for many nuclei from
fusion measurements through CC analyses (see, {\it e.g.},
Refs.~\cite{Leigh1995,Rumin1999}). Although this procedure has inherent
limitations (the extracted results depend sensitively on the accuracy
of the experimental data and on uncertainties that are not always well
constrained, besides the derived observables exhibiting some degree
of model dependence),
reliability of this approach has been systematically validated across a
wide range of nuclear systems.

A more sensitive representation of fusion data
for studying channel coupling effects
is the fusion barrier distribution ($\mathcal{D}$), proposed by Rowley
\textit{et al.}~\cite{Rowley1991}. It is derived from the measured fusion
excitation function using the double-derivative (DD) method applied to
$E\sigma_{\mathrm{fus}}$ ($E$ is energy available in the center-of-mass
frame of reference), typically evaluated using a point-difference formula.
However, the DD method is highly sensitive to the chosen step size
$\Delta E$) and
exhibits poor accuracy at higher energies ~\cite{Dasgupta1998}, limiting
the structural information that can be reliably extracted. In the
literature, several alternative approaches for determining the $\mathcal{D}$
have been proposed (see Refs.~\cite{Chandra2025PLB,Chandra2025PRC}).
Among these, the Gaussian analytic recipe~\cite{Jiang2022} is particularly
robust, being independent of step size, resilient to large uncertainties
in $\sigma_{\mathrm{fus}}$ and capable of providing a complete,
{\it well-defined} spectrum across the barrier
~\cite{Chandra2025PLB,Chandra2025PRC,Jiang2022}. 
This makes it a powerful method for investigating nuclear
structure through fusion dynamics. For instance, recently the ground-state
(g.s.) deformation parameters of the $sd$-shell nucleus $^{28}$Si
~\cite{Chandra2025PLB} and the rare-earth nucleus $^{154}$Sm
~\cite{Chandra2026prc} were successfully extracted using this approach,
utilizing fusion data, respectively, from the $^{28}$Si+$^{144}$Sm
and $^{16}$O+$^{154}$Sm systems, in conjunction with Bayesian analyses.

In CC calculations, model parameters are often adjusted to reproduce
experimental data for a specific system. However, such system-dependent
tuning can sometimes yield parameter values with limited physical
justification. Moreover, deformation parameters extracted in this manner
may not provide a consistent description when applied to different
projectile-target combinations or to complementary reaction channels.
For example, a study of the $^{28}$Si+$^{90}$Zr system by Gupta
\textit{et al.}~\cite{Gupta2023} suggested an oblate shape for $^{28}$Si
based on the extracted $\beta_2$ values and reported a vanishingly small
$\beta_4$ deformation. In contrast, a recent study of the
$^{28}$Si+$^{144}$Sm system by Kumar and Nath ~\cite{Chandra2025PLB}
treated $^{28}$Si as either a rotor or a vibrator and, based on their
results, could not distinguish between prolate and oblate shapes or
confirm a purely rotational / vibrational character for $^{28}$Si. The
authors of Ref. \cite{Chandra2025PLB} also suggested a comparatively
large $\beta_4$ value, in clear contradiction to the findings of Gupta
\textit{et al.} Similarly, a large $\beta_4$ value for $^{28}$Si was
reported from the study of the $^{28}$Si+$^{92}$Zr system~\cite{Kaur2018}.
These ambiguities highlight the limitations of parameter tuning restricted
to individual systems, which might lead to inconsistent or even unphysical
results. Therefore, it is essential to constrain arbitrary parameter
variations and to seek a simultaneous reproduction of experimental data
across multiple systems. Examining the consistency and robustness of
extracted deformation parameters for a given nucleus is crucial for
validating the reliability of the adopted methodology.

In this context, we aim to 
investigate the \emph{shape} of $^{140}$Ce,
a well-known vibrational nucleus, and to examine consistency of the
results when constrained by multiple reaction systems. For this
nucleus, different experimental probes yielded markedly different
deformation parameters. A recent study based on the isovector giant
dipole resonance (IVGDR) \cite{Kleemann2024} predicted a very small
quadrupole deformation for $^{140}$Ce, indicating an almost spherical
shape with $\beta_2 = 0.0030(15)$, whereas Ref.~\cite{Pritychenko2016}
reported a significantly larger value of $\beta_2 = 0.1018(25)$.
In contrast, Montagnoli \textit{et al.}~\cite{Montagnoli2006}
suggested an even larger deformation, $\beta_2 \approx 0.15$,
based on fusion measurements. It is important
to note here that deformation, in the context of a vibrational
nucleus, does not refer to a static and rigid non-spherical shape.
It rather implies a dynamic feature of the nucleus, characterized
by collective oscillations of its surface around an equilibrium
spherical shape.

As a first step towards this objective, we analyze newly measured fusion
data for the $^{16}$O+$^{140}$Ce system together with existing experimental
$\sigma_\text{fus}$ for the $^{36}$S+$^{140}$Ce system~\cite{Montagnoli2006}.
Both reactions are particularly favourable to probe deformation of the target
nucleus, as the projectiles, {\it viz.}, $^{16}$O and $^{36}$S, are spherical
and free from the complexity to be encountered for deformed projectiles
in CC calculations. These two systems also do not have any positive $Q$-value
neutron transfer (PQNT) channels. Given that the interplay between
inelastic excitations and transfer couplings remains an open problem and
cannot yet be treated fully quantum mechanically within CC formalisms
~\cite{Cook2023}, the $^{16}$O+$^{140}$Ce and $^{36}$S+$^{140}$Ce systems
provide reliable benchmarks for extracting
 the dynamic deformation parameters
for $^{140}$Ce. To place the results in a broader context, we further
consider existing experimental fusion excitation function for the
$^{28}$Si+$^{140}$Ce system~\cite{Chandra2025PRC}, which is considerably
more complex due to the intrinsic deformation of the projectile ($^{28}$Si)
and the presence of PQNT channels, and therefore does not provide
definitive constraints on the deformation parameters of $^{140}$Ce
by itself. 

We extract deformation parameters
of $^{140}$Ce from the fusion data for $^{16}$O+$^{140}$Ce and
$^{36}$S+$^{140}$Ce systems and subsequently test their validity
against the $^{28}$Si+$^{140}$Ce data. It should also be noted
that the $^{36}$S+$^{140}$Ce system was previously studied through
extraction of the $\mathcal{D}$ using the conventional DD method,
which is known to suffer from large uncertainties at higher energies,
as previously discussed. In the present work, we have
derived $\mathcal{D}$ for this system also, using the Gaussian analytic
recipe, to have a clearer understanding of the fusion dynamics.

\begin{figure}[hbt!]
   \centering
   \includegraphics[width=0.80\linewidth]{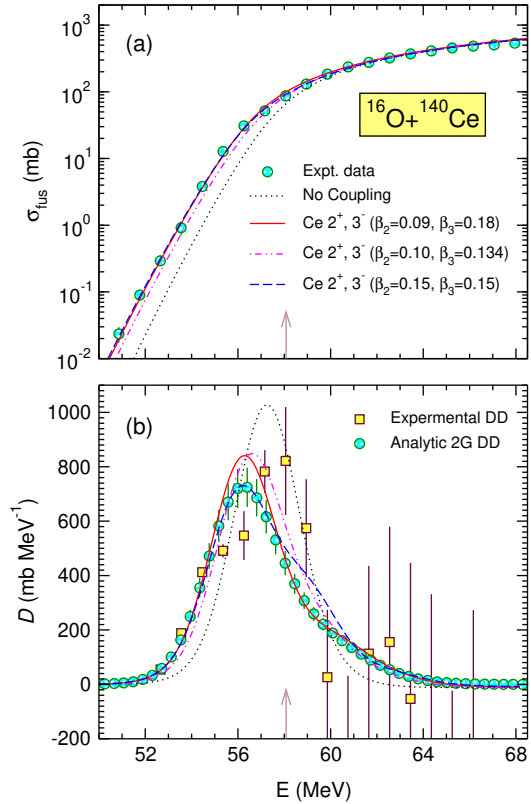}    
   \caption{\label{fig:fig01} Measured and calculated (a) fusion excitation
   function and (b) barrier distribution for $^{16}$O+$^{140}$Ce. Experimental
   data (squares) and Gaussian analytic results (circles) are compared with
   CC calculations ($\Delta E\simeq1.8$~MeV) in the lower panel. The Bass
   barrier is marked by an upward arrow in each panel.}
\end{figure}

The fusion measurement for $^{16}$O+$^{140}$Ce
has been performed using the Heavy Ion Reaction
Analyzer (HIRA)~\cite{Sinha1994} with a pulsed $^{16}$O beam (4~$\mu$s
pulse separation) from the 15UD Pelletron. An isotopically enriched
$^{140}$Ce ($\sim$200~$\mu$g/cm$^2$) film, with thin carbon backing
and capping layers ($\sim$20~$\mu$g/cm$^2$ and $\sim$5~$\mu$g/cm$^2$,
respectively)~\cite{RBVac2021} has been used as the target. Beam energy
($E_\text{lab}$) has been varied in the range 57 to 76 MeV in steps
of 1 MeV, from $\simeq$12.4\% below to $\simeq$17\% above the Bass
barrier \cite{Bass1980}. Further experimental details can be found
in Refs.~\cite{Biswas2024,Biswas2026}.
The measured fusion excitation function for the $^{16}$O+$^{140}$Ce
system is shown in Fig.~\ref{fig:fig01}(a). The corresponding $\mathcal{D}$,
shown in Fig.~\ref{fig:fig01}(b), has been derived using
(i) the DD method proposed by Rowley \textit{et al.}~\cite{Rowley1991},
applied directly to the experimental data, and (ii) the Gaussian analytic
recipe prescribed by Jiang \textit{et al.} \cite{Jiang2022}.
It is clear from Fig.~\ref{fig:fig01}(b) that the $\mathcal{D}$
derived from measured $\sigma_\text{fus}$ by the DD method suffers
from large uncertainties at higher values of $E$.

\begin{figure}[hbt!]
   \centering
   \includegraphics[width=0.80\linewidth]{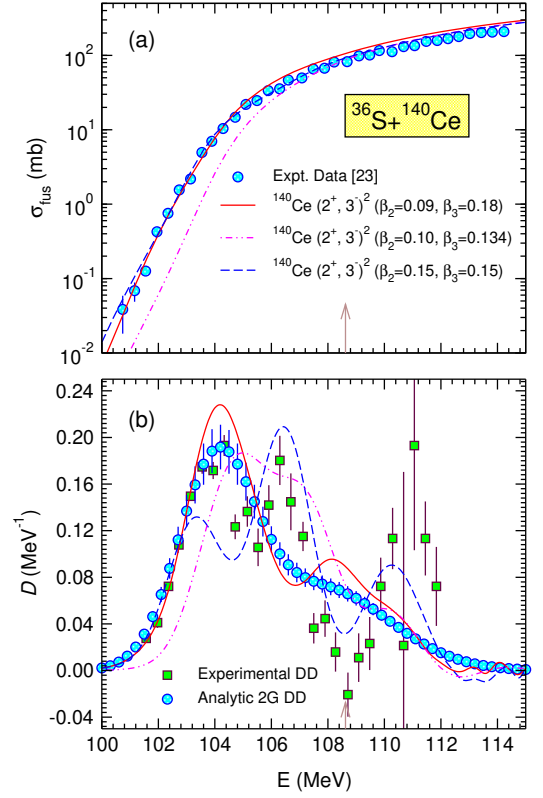}    
   \caption{\label{fig:fig02} Measured and calculated (a) fusion excitation
   function and (b) barrier distribution (normalized to unity) for
   $^{36}$S+$^{140}$Ce~\cite{Montagnoli2006}. Experimental data (squares)
   and Gaussian analytic results (circles) are compared with CC calculations
   ($\Delta E\simeq1.6$~MeV) in the lower panel. The Bass barrier, in each
   panel, is indicated by an upward arrow.}
\end{figure}

The optimized Gaussian parameters for the two reactions are listed
in Table~\ref{table:Multi_Gaussian}. For $^{16}$O+$^{140}$Ce, two
Gaussian components are sufficient. Transitioning from a single-Gaussian
(1G) to a two-Gaussian (2G) fit changes the reduced $\chi^{2}$ from 0.8897
to 0.2298 without significantly altering the overall shape of $\mathcal{D}$.
The statistical sensitivity of the fits are
discussed in the $\texttt{Appendix}$.
A similar analysis has been carried out for $^{36}$S+$^{140}$Ce,
which is depicted in Fig. \ref{fig:fig02}.
As shown in Table~\ref{table:Multi_Gaussian}, the reduced $\chi^{2}$
decreases from 2.9575 (1G) to 2.3865 (2G) and further to 1.5914 (3G).
However, in the 3G fit, the standard deviation of the second Gaussian
becomes negligible ($\texttt{W}_2 = 0.0019$~MeV), confirming that two
Gaussian components adequately describe $\mathcal{D}$ for this system
as well.
Notably for $^{36}$S+$^{140}$Ce, the conventional DD method results
in a three-peak-like structure of the $\mathcal{D}$ (Fig.~\ref{fig:fig02}(b)).
In contrast, the Gaussian analytic recipe produces a major peak at
$E \simeq 104$ MeV and a broad peak-like structure at higher energies.

To further interpret these results, CC calculations have been performed
using \textsc{ccfull}~\cite{Hagino1999} to reproduce the measured fusion
excitation functions and the analytical $\mathcal{D}$s for both systems.
The nuclear potential has been assumed to have a Woods-Saxon form. For
$^{16}$O+$^{140}$Ce, the reduced radius ($r_\circ$)
has been slightly adjusted from 1.20 fm (used for
$^{16}$O+$^{142}$Ce \cite{Biswas_fusion23}) to 1.185 fm, while
keeping other potential parameters unchanged.

\begin{table*}[hbt!]
\centering
\caption{\label{table:Multi_Gaussian}Results of multi-Gaussian
fits for the systems $^{16}$O+$^{140}$Ce and $^{36}$S+$^{140}$Ce. 
$\mathbb{L}$ is the number of Gaussian components; $\texttt{V}_i$,
$\texttt{W}_i$, and $\omega_i$ denote centroid, standard deviation
and weight of the $i$-th component, respectively. $\texttt{R}$ is
the barrier radius. $\chi_0^2$ and $\chi^2$ are the usual measures
of goodness of fit with
$\chi^2 (\mathbb{L} \textrm{G}) = \frac{N}{N-\nu}\ \chi_{0}^{2} (\mathbb{L} \textrm{G})$,
where $N$ and $\nu$ stand for the number of experimental data
points and the degrees of freedom in the fitting procedure,
respectively.}
\begin{tabular}{cccccccccccccc}
\hline
Systems & $\mathbb{L}\textsc{G}$ & $\chi_{0}^2$($\mathbb{L}$\textsc{G}) &  $\chi^2$($\mathbb{L}$\textsc{G}) & \texttt{R} & $\omega_1$ & \texttt{V}$_1$ & \texttt{W}$_1$  & $\omega_2$ & \texttt{V}$_2$  & \texttt{W}$_2$ & $\omega_3$ & \texttt{V}$_3$  & \texttt{W}$_3$ \\ 
 &  &  &  & (fm) &  & (MeV) & (MeV) &  & (MeV) & (MeV)  &  & (MeV) & (MeV) \\
\hline
\multirow{2}{*}{$^{16}$O+$^{140}$Ce} & 1G & 0.7562 & 0.8897 & 10.70 & 1.00 & 57.01 & 1.81 &  &  & &  &  & \\
 & 2G & 0.1608 & 0.2298 & 10.56 & 0.57 & 56.03 & 1.21 &0.43 & 58.20 &2.41 &  &  &\\
 
\hline
\multirow{2}{*}{$^{36}$S+$^{140}$Ce} & 1G & 2.704 &2.9575 & 8.63 & 1.0 &104.51 &1.37  &  &  & &  &  & \\
 & 2G &1.9774  & 2.3865 &9.56 &0.56 &104.03 &1.17 & 0.44& 107.75& 2.39 &  &  & \\
  & 3G & 1.1822 & 1.5914 & 9.05 & 0.02&101.67  & 0.044& 0.14 & 103.14 &0.0019  & 0.84 &105.35  &1.647 \\
\hline
\end{tabular}
\end{table*}


The results of CC calculations are compared with the measured
excitation functions and the $\mathcal{D}$s in Fig.
~\ref{fig:fig01}. As expected, the no-coupling calculation
(black dotted curve in panel (a))
reproduces the experimental $\sigma_\mathrm{fus}$ only above the
barrier, while significantly underestimating the data below it.
The calculations
have been improved by including only the internal degrees of freedom
of the target nucleus, as inclusion of the first excited state of
$^{16}$O is known to cause overprediction of $\sigma_\text{fus}$
above the barrier and merely shifts the theoretical $\mathcal{D}$
by a constant amount~\cite{Hagino1997}.
 Here,
$^{140}$Ce has been modeled as a purely vibrational
nucleus, incorporating one-phonon quadrupole ($2^+$) and octupole
($3^-$) excitations, following the approach for $^{16}$O+$^{144}$Sm
~\cite{Hagino1997}. To determine the deformation parameters of $^{140}$Ce,
we have explored a broad parameter space for $\beta_2$ and $\beta_3$,
systematically varying both from 0.00 to 0.60 in steps of 0.01,
using a coupling Hamiltonian radius parameter of 1.20 fm.

For $^{36}$S+$^{140}$Ce, we have used potential parameters, projectile
and target couplings, as prescribed in the Ref. \cite{Montagnoli2006}.
Similar to the case of $^{16}$O+$^{140}$Ce, $\beta_2$ and $\beta_3$
have been varied between 0.00 and 0.60 with an increment of 0.01 for
both deformation parameters.

Next, quality of the fit between the 2G analytic $\mathcal{D}$ and
the theoretical $\mathcal{D}$ obtained from \textsc{ccfull}, for both
systems, has been quantified using the relation
\begin{equation}\label{eq:eq01}
\chi^2_{\textrm{fus}}(\beta_2, \beta_3) = \frac{1}{N} \sum_{i=1}^{N} \left[ \frac{\texttt{D}_{\textrm{fus}_i}^{\text{2G}} -\texttt{D}_{\textrm{fus}_i}^\text{CC}}{\delta \texttt{D}_{\textrm{fus}_i}^\text{2G}} \right]^2\ .
\end{equation}
\noindent
Here
$\delta \texttt{D}_{\textrm{fus}_i}^{\text{2G}}$ denote the uncertainties
in the analytic barrier distribution, which are assumed to be 10\%.  
The $\chi^2_{\textrm{fus}}(\beta_2, \beta_3)$ analysis has been
performed over the energy ranges of 51 \textendash 65 MeV and
100 \textendash 112.5 MeV for $^{16}$O+$^{140}$Ce and $^{36}$S+$^{140}$Ce,
respectively.

\begin{figure*}[hbt!]
    \centering
    \includegraphics[width=0.75\linewidth]{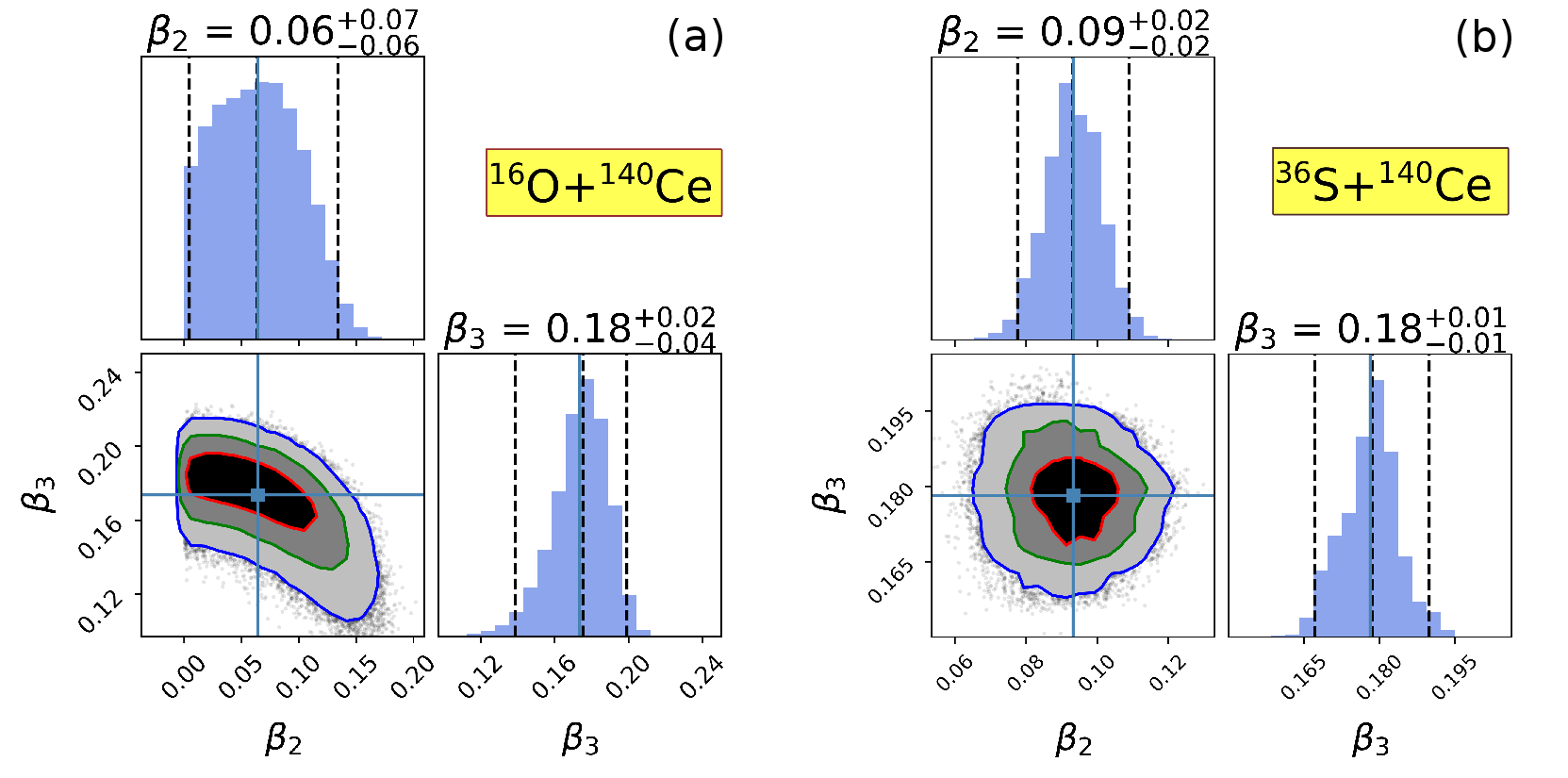}    
    \caption{\label{fig:fig03}
    Posterior probability distributions of the deformation parameters
    $\beta_2$ and $\beta_3$ obtained from the Bayesian analysis of the
    systems (a) $^{16}$O+$^{140}$Ce and (b) $^{36}$S+$^{140}$Ce,
    for the nucleus $^{140}$Ce. The one-, two- and three-sigma confidence
    contours are displayed. The quoted uncertainties correspond to the
    95\% (two-sigma) confidence intervals of the estimated deformation
    parameters.}
\end{figure*}

Subsequently, a Bayesian analysis within the Markov Chain Monte
Carlo (MCMC) framework has been performed to determine $\beta_2$
and $\beta_3$ along with their associated uncertainties in each system.
A detailed description of the methodology is provided in 
Ref.~\cite{Chandra2025PLB}. The posterior probability distributions
yield mean values of $\beta_2 = 0.06^{+0.07}_{-0.06}$ and
$\beta_3 = 0.18^{+0.02}_{-0.04}$ for the $^{16}$O+$^{140}$Ce system,
exhibiting a moderate correlation $\sim$ 0.62. For $^{36}$S+$^{140}$Ce,
the corresponding values are $\beta_2 = 0.09^{+0.02}_{-0.02}$ and
$\beta_3 = 0.18^{+0.01}_{-0.01}$, with only a weak correlation of
about 0.07. The corresponding corner plots of the posterior probability
distributions are presented in Fig. \ref{fig:fig03}. Next, these
two independent estimates of the deformation parameters are combined
by application of the iNdependent Bayesian Model Averaging (NBMA)
technique, described in Ref.~\cite{Chandra2026prc}. Thus, $\beta_2$
and $\beta_3$ of $^{140}$Ce are optimally inferred to be
$0.09^{+0.03}_{-0.03}$ and $0.18^{+0.02}_{-0.02}$ from this work.
The value of quadrupole deformation, obtained from the present study,
is quite close to the one reported by Pritychenko {\it et al}
($\beta_2 = 0.10$) \cite{Pritychenko2016}, while the value of the
octupole deformation is larger than the ones reported earlier
\cite{Montagnoli2006,Kibedi2002}.

CC calculations incorporating the optimized $\beta_2$ and $\beta_3$
of $^{140}$Ce have been compared with the experimental fusion excitation
functions and the corresponding $\mathcal{D}$s for $^{16}$O+$^{140}$Ce
and $^{36}$S+$^{140}$Ce in Fig.~\ref{fig:fig01} and Fig.~\ref{fig:fig02},
respectively. The data are well reproduced by the calculations with the
theoretical and the experimental peak positions in the $\mathcal{D}$s
in close agreement.

For comparison, CC calculations using previously reported deformation
parameters of $^{140}$Ce are also shown in Fig.~\ref{fig:fig01} and
Fig.~\ref{fig:fig02}. When adopting the deformations ($\beta_2 = 0.10$
and $\beta_3 = 0.134$) from Refs.~\cite{Pritychenko2016,Kibedi2002},
the calculated excitation function for $^{16}$O+$^{140}$Ce
($^{36}$S+$^{140}$Ce) slightly (significantly) underpredict the measured
$\sigma_\text{fus}$. The corresponding theoretical $\mathcal{D}$s show
clear deviations from analytical ones, both in barrier position and shape.
Use of the deformation parameters ($\beta_2 = 0.15$ and $\beta_3 = 0.15$)
from Ref.~\cite{Montagnoli2006} worsens the agreement between the
analytical and the calculated $\mathcal{D}$s, with respect to the
results based on the currently determined values of $\beta_2$ and
$\beta_3$, for $^{16}$O+$^{140}$Ce. In case of $^{36}$S+$^{140}$Ce,
the calculation produces an additional, high-energy peak that is absent
in the analytic $\mathcal{D}$.


\begin{figure}[htb!]
   \centering
   \includegraphics[width=0.87\linewidth]{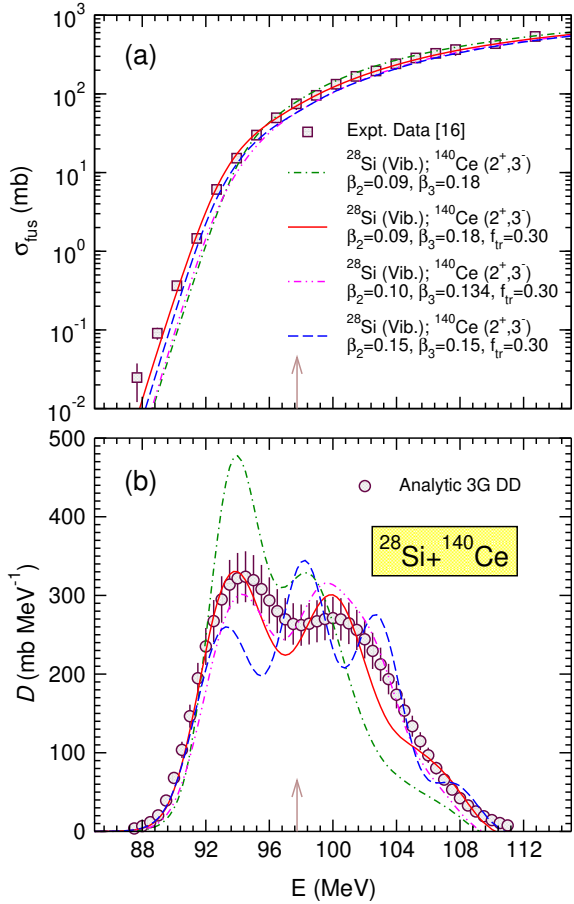}    
   \caption{\label{fig:fig04} Measured and calculated (a) fusion excitation
   function and (b) barrier distribution for $^{28}$Si+$^{140}$Ce
   \cite{Chandra2025PRC}. Gaussian analytic results (circles) are compared
   with CC calculations ($\Delta E\simeq2.5$~MeV) in the lower panel.
   The upward arrow, in each panel, indicates the Bass barrier.}
\end{figure}

Subsequently, CC calculations have been performed for $^{28}$Si+$^{140}$Ce,
using similar potential parameters but a revised coupling scheme relative
to our earlier work~\cite{Chandra2025PRC}. The present scheme includes
one-phonon $2^+$ and $3^-$ excitations of $^{140}$Ce and the first
excited state of $^{28}$Si, treated as both a vibrator and a rotor.
Deformation parameters for $^{140}$Ce have been taken from the present
analysis and the same for $^{28}$Si from Ref.~\cite{Chandra2025PLB}.
The results are shown in Fig.~\ref{fig:fig04}, where $^{28}$Si is treated
first as a perfect vibrator (green dash-dotted curve). These couplings
alone are unable to reproduce the measured fusion excitation function
and the 3G analytical $\mathcal{D}$.
It is important to note that the
pair transfer channel has not been included in this initial CC calculation,
despite the presence of two-neutron ($2n$) transfer channel with g.s. to
g.s. transfer $Q$-value of 2.43 MeV.
PQNT  channels are known to significantly influence fusion dynamics in
many systems. Therefore, we have manually varied the pair-transfer
strength ($f_\text{tr}$) in the CC calculations and have found that
$f_\text{tr} = 0.30$ fm provides good reproduction of both
$\sigma_\text{fus}$ and analytical $\mathcal{D}$ (red solid line in
Fig. \ref{fig:fig04}. A pronounced enhancement of sub-barrier cross
sections in $^{28}$Si+$^{140}$Ce due to inclusion of the PQNT channel
in CC calculation is noted, in contrast to our earlier conclusion
(see Ref. \cite{Chandra2025PRC}) where only a minimal enhancement was
observed. Furthermore, inclusion of the transfer coupling is found to
be essential to reproduce the analytical $\mathcal{D}$, as clearly
shown in Fig.~\ref{fig:fig04}(b). To further investigate the influence
of projectile structure, $^{28}$Si has also been treated as a rigid
rotor. The results (not shown in the figure) reveal only marginal
changes in $\sigma_\text{fus}$ and $\mathcal{D}$, regardless of whether
$^{28}$Si is assumed to have an oblate or a prolate.
In all these calculations, we have used $\beta_2 = 0.28$ for $^{28}$Si,
obtained from the analysis of fusion data of $^{28}$Si+$^{144}$Sm
~\cite{Chandra2025PLB}. Additionally, when the deformation parameter(s)
extracted from quasielastic scattering for the same system
~\cite{Chandra2025PLB} are used in the CC calculations, the resulting
$\mathcal{D}$ shows poorer agreement with analytical
$\mathcal{D}$ (not shown in the figure).

Next, we have compared the present results with those obtained using
previously reported deformation parameters of $^{140}$Ce.
When adopting the deformations from Refs.~\cite{Pritychenko2016,Kibedi2002},
the calculated excitation functions for $^{28}$Si+$^{140}$Ce largely
underpredict the measured $\sigma_\text{fus}$ (magenta dash-double-dotted
curve in Fig. \ref{fig:fig04}).
However, the corresponding analytical $\mathcal{D}$ is comparatively
better reproduced, likely because the barrier distribution is more
sensitive to the shape of the excitation function than its absolute
magnitude.  Use of
deformation parameters of $^{140}$Ce from Ref.
~\cite{Montagnoli2006} yields better matching with the experimental
excitation function (blue dashed curve) but leads to a theoretical
$\mathcal{D}$ with a distinctly different shape from the analytical
one.  We emphasize that
inclusion of the previously determined
deformation parameter(s) of $^{28}$Si \cite{Chandra2025PLB} and the
presently extracted $\beta_2$ and $\beta_3$ of $^{140}$Ce in the CC
calculation reproduces both the $\sigma_\text{fus}$
and the $\mathcal{D}$
for this system, indicating robustness of the adopted coupling schemes.

In summary, fusion cross sections for the $^{16}$O+$^{140}$Ce
system have been measured down to $\sim 12.4\%$ below the Bass
barrier and barrier distribution has been derived using both the
conventional DD method and the Gaussian analytic recipe. Application
of the Gaussian analytic method to $^{36}$S+$^{140}$Ce has yielded
a well-defined $\mathcal{D}$ over the full energy range. The three-peak
structure reported for this system earlier \cite{Montagnoli2006} has
not been found in the Gaussian analytic results, pointing to the
possibility of its appearance because of the limitations of the DD
approach. Coupled-channels calculations have been performed for both
the systems, to determine the dynamic quadrupole and octupole
deformation parameters of $^{140}$Ce
by adopting the $\chi^2$-minimization technique. Bayesian uncertainty
analysis has been carried out and optimal values of $\beta_2$ and
$\beta_3$ have been obtained using NBMA. Further, extracted deformation
parameters have been used in the $^{28}$Si+$^{140}$Ce system. These
parameters reproduce both the fusion excitation function and the
analytic $\mathcal{D}$, when previously determined deformation of
$^{28}$Si \cite{Chandra2025PLB} is incorporated in the CC calculations.
Whether $^{28}$Si is treated as a vibrator or a rotor has a negligible
effect on the results. Inclusion of transfer of a pair of neutrons in
the CC calculations causes a noticeable enhancement of the sub-barrier
fusion cross section in $^{28}$Si+$^{140}$Ce, emphasizing the significant
role of the PQNT channel in this system.

Overall this work presents a comprehensive investigation of nuclear
deformation effects in heavy-ion fusion reactions by combining
experimental data, a Gaussian analytic recipe for deriving the
$\mathcal{D}$ and CC calculations. For the first time, we demonstrate
the consistency and robustness of the extracted deformation parameters
of a nucleus across reactions induced by three distinct projectiles.
More such studies, aiming for simultaneous reproduction of experimental
observables across multiple systems, would be quite effective
in constraining extracted deformation parameters.

The authors thank the Pelletron crew of IUAC for providing
pulsed beam with requisite specifications and the staff of
the IUAC Target Laboratory for excellent support in preparing
the enriched target films. The authors acknowledge the National
Supercomputing Mission (NSM) for providing computing resources
of `PARAM Rudra' installed at the Inter-University Accelerator
Centre, Aruna Asaf Ali Marg, New Delhi 110067, which is
implemented by the Centre for Development of Advanced Computing
(C-DAC) and supported by the Ministry of Electronics and
Information Technology (MeitY) and the Department of Science
and Technology (DST), Government of India.

\appendix
\input{Appendix}
\end{document}

%% file: Appendix.tex
\begin{center}
{
{\bf{\Large{\textsf{Appendix}}}}\\~\\
}
\end{center}
 The increase of the number of Gaussians in the fitting
procedure leads to a reduction in $\chi_0^2$. A pertinent question to
ask is whether the extra Gaussian indicates a physically significant
peak in the fusion barrier distribution ($\mathcal{D}$). The reduced
$\chi_0^2$ (denoted by $\chi^2$ in the article) is often taken as a
measure of the goodness of fit. However, it may not always penalize
the complexity of the model (\textit{i.e.}, inclusion of more
Gaussian components and thereby more fit parameters) strongly
enough.

In order to examine the possibility of overfitting in the Gaussian
analytic recipe for $\mathcal{D}$ \cite{Jiang2022}, we have carried
out several statistical tests. In particular, the
$F$-test~\cite{Bevington1993}, as well as the Akaike Information
Criterion (AIC)~\cite{Akaike1974,Burnham2002} and the Bayesian
Information Criterion (BIC)~\cite{Schwarz1978,Robert1995},
have been used to evaluate the quality of the fits and to
compare the relative performance of the one-Gaussian (1G)
and the two-Gaussian (2G) models for the $^{16}$O+$^{140}$Ce
system. These complementary statistical measures provide
a more robust basis for assessing whether the inclusion
of additional parameters is justified.

We have used the standard prescription for evaluating the
$F$-test \cite{Bevington1993}, which is given by

\begin{equation}
F = \frac{\left(\chi^2_\text{1G} - \chi^2_\text{2G}\right)/\nu_1}
{\chi^2_\text{2G}/\nu_2};
\qquad
\nu_1 = k_\text{2G}- k_\text{1G}, \quad \nu_2 = N - k_\text{2G}.
\end{equation}

\noindent
Here, $\chi^2_{j}$ ($j = \text{1G}$, 2G) are defined as
\begin{equation}
\chi^2_j = \sum_{i=1}^{N} \left( \frac{\sigma_{\mathrm{fus}_i}^{\mathrm{expt.}} - \sigma_{\mathrm{fus}_i,j}^\mathrm{model}}{\delta \sigma_{\mathrm{fus}_i}^\mathrm{expt.}} \right)^2.
\end{equation}

\noindent
The quantities $\sigma_{\mathrm{fus}_i}^\text{expt.}$,
$\sigma_{\mathrm{fus}_i,j}^{\mathrm{model}}$ and
$\delta\sigma_{\mathrm{fus}_i}^\text{expt.}$ denote the experimental
fusion cross sections, the corresponding analytic fusion cross sections
predicted by the 1G and 2G models and the experimental uncertainties,
respectively. $N$ is the number of data points, while $k_\text{1G}$
and $k_\text{2G}$ represent the number of free parameters in the 1G
and 2G models, respectively.

For the $^{16}$O+$^{140}$Ce system, the results of 1G and 2G analyses
(with $N = 20$, $k_\text{1G} = 3$ and $k_\text{2G} = 6$) are given by
\begin{align}
\chi^2_{\mathrm{1G}} &= 15.124, \\
\chi^2_{\mathrm{2G}} &= 3.216.
\end{align}

The $F$-statistic is found to be
\begin{equation}
F \approx 17.28.
\end{equation}

To assess whether the 1G or the 2G model is preferable, the computed
$F$ is compared with the critical value of the $F$-distribution at
the 5\% significance level. For $\nu_1 = 3$ and $\nu_2 = 14$, the
critical value is $F_{0.05}(3,14) \approx 3.34$~\cite{OttLongnecker2010}.
Since the calculated $F$-statistic for the system $^{16}$O+$^{140}$Ce 
is significantly larger than $F_{0.05}$, we conclude that the 2G model
provides a statistically significant improvement over the 1G model.

In addition to the $F$-test, we have also employed the AIC and the
BIC to assess the statistical significance of the two models.
These statistical metrics are defined as
\begin{subequations} \label{eq:AIC_BIC_I}
  \begin{alignat}{1}
    & \mathrm{AIC}_j = -2 \ln \mathcal{L}^\text{max}_j + 2k_j \label{eq:AIC_I}, \\
    & \mathrm{BIC}_j = -2 \ln \mathcal{L}^\text{max}_j + k_j \ln N. \label{eq:BIC_I}
  \end{alignat}
\end{subequations}

\noindent
where $\mathcal{L}^\text{max}_j$ are the maximum likelihood
for the 1G and 2G models. For Gaussian-distributed uncertainties,
the log-likelihood is related to the chi-square as
\begin{equation}
-2 \ln \mathcal{L}^\text{max}_j = \chi^2_j + \text{constant}.
\end{equation}

Since the constant term is independent of the model parameters
and therefore does not affect model comparison, Eq. \ref{eq:AIC_I}
and \ref{eq:BIC_I} reduce to
\begin{subequations} \label{eq:AIC_BIC_II}
  \begin{alignat}{1}
    & \mathrm{AIC}_j = \chi^2_j + 2k_j \label{eq:AIC_II}, \\ 
    & \mathrm{BIC}_j = \chi^2_j + k_j \ln N \label{eq:BIC_II},
  \end{alignat}
\end{subequations}

\noindent
respectively.

The calculated values are
\begin{align}
\mathrm{AIC}_{\mathrm{1G}} &= 21.124, \quad & \mathrm{AIC}_{\mathrm{2G}} &= 15.216, \\
\mathrm{BIC}_{\mathrm{1G}} &= 24.111, \quad & \mathrm{BIC}_{\mathrm{2G}} &= 21.190.
\end{align}

We note that both the AIC and the BIC decrease, going from 1G
to 2G model. Both of these quantities support the 2G model as
providing a better balance between goodness of fit and model
complexity. Based on all statistical indicators, the 2G model
is consistently favored over the 1G model in the present case.
This suggests that the additional Gaussian component captures
a genuine feature of the $\mathcal{D}$ for $^{16}$O+$^{140}$Ce,
rather than simply overfitting the data.

Finally, we must examine whether the extra Gaussian in the
2G model is physically meaningful. It can be noted that our
coupled-channels analyses, across three systems, consistently
correlate the computed fusion barrier distributions with known
structural features of the collision partners.